\documentstyle[prl,aps,preprint]{revtex}
\tightenlines

\begin{document}

\draft
\preprint{\vbox{
\hbox{UMD-PP-00-099}}}

\title{About the Overcompleteness of Coherent State Systems with a
 Line Bundles Viewpoint}

\author{A. I. Shimabukuro\footnote{e-mail:shima@ift.unesp.br}}
\address{Instituto de F\'isica Te\'orica, Univeridade Estadual Paulista,\\
Rua Pamplona 145, 01405-900, S\~ao Paulo, S\~ao Paulo, Brasil
}

\date{February, 2001}

\maketitle
\begin{abstract}
{Standart Coherent State Systems have an analysis based on 
lattices (von Neumanns's lattices) in terms of wich they are
 classified, looking at the size of the minimun cell, by: 
complete, overcomplete and not complete. In this work we 
analize overcomplete systems with a  geometrical 
viewpoint (holomorphic line-budles). We apply the method to evaluate 
the degeneracy of the lowest Landau level.}     

\end{abstract}

\section{Introduction}
In this work, we are going to investigate the question of overcompleteness
of a Coherent State System using the framework of line bundles.
By the von Neumann construction (von Neumann lattices) we know
that complete systems are related to a certain type of lattice (what we have 
called complete lattice) and overcomplete systems are related to another one 
(overcomplete lattice). Analyzing the representation of the Weyl-Heisenberg 
group induced by characters on these lattices, we associate to a complete
system a theta function and with the overcomplete system a set of theta 
functions with characteristic.

Since theta functions are the sections of line bundles over a complex torus, 
and since these line bundles can be seen as the holomorphic quantization of 
a classical system, we have a direct interpretation of the
overcompleteness 
of the system: an overcomplete system (associated with an overcomplete
lattice) is a set of quantizations of this classical system. This is
expected since to have overcomplete systems we've had to consider more than 
one state by Planck cell. We apply this result to analyze the degeneracy
of the 
lowest Landau level by using the Riemann-Roch theorem in the appropriate line
bundle.

In the section 2 we present briefly the question of the overcompleteness. In 
section 3 we remind some aspects of representation of the Weyl-Heisenberg group
induced by characters on the lattice and how it's related to theta functions.
In the section 4 we associate this result to holomorphic quantizations of the 
classical system. And in the last section we make the application of the method to Landau levels.

\section{The Overcompleteness of C.S.S.}

Let $w$\ be the Weyl-Heisenberg algebra generated by $\left\{ \hat{
p},\hat{q},I\right\} $, with the usual commutation relations:

\begin{eqnarray}
\left[ \hat{p},\hat{q}\right]  &=&i\hbar I,  \nonumber \\
\left[ \hat{q},I\right]  &=&\left[ \hat{p},I\right] =0  \label{1}
\end{eqnarray}

We have the representation in terms of the creation and annihilation
operators: $a=({\hat{q}+i\hat{p}})/{\sqrt{2\hbar }}$, $a^{+}=({\hat{q}
-i\hat{p}})/{\sqrt{2\hbar }}$, such that $\left[ a,a^{+}\right] =I$.

We will use different notations to represent an element $x$ of $w$:

\begin{equation}
x=itI+\frac{i}{\hbar }\left( P\hat{q}-Q\hat{p}\right) =itI+\left( \alpha
a^{+}-\bar{\alpha}a\right)   \label{2}
\end{equation}
for \ $P,Q,t\in {\bf R}$ and $\alpha =\left( 2\hbar \right) ^{-1/2}(Q+iP)$.

We represent $x$ also by $x=( t,\vec{v})$
or $x=\left(t,\vec{\alpha}\right)$ where $\vec{\alpha}=\alpha
a^{+}-\bar{\alpha}a$, $\vec v=\left( P,Q\right) $ is a
point in the $V\times V$ plane and $t$ is the central term of the algebra.

The elements of the Weyl-Heisenberg Group $W$ is obtained by the
exponentiation map:

$\exp(x)=\exp(itI)\exp{\left( \alpha a^{+}-\bar{\alpha}a \right)}$,\\ 
and because of the commutation relations, we have:\\
\begin{equation}
\exp{A} \exp{B} = \exp(\frac{1}{2}[A,B])\exp(A+B)   \label{3}
\end{equation}
for $A,B \in w$.

We use also to represent the above equation as:\\
\begin{equation}
(t,\vec v).(t^{\prime},\vec {v}^\prime)=\left( t+t^{\prime}+\frac{1}{2}
B(\vec v,\vec{v}^\prime),\vec v+\vec{v}^\prime\right)   \label{4}
\end{equation}
where $B:V\times V \rightarrow \bf{R}$ is the alternating bilinear form defined by 
$B(\vec{v},\vec{v}^\prime)=\left[ \vec{v},\vec{v}^\prime \right]$.

Observe that we have used (abusing on notation) the same symbols 
$(t,\vec v)$ to indicate both, elements of the algebra $w$ and the group
$W$.

Let $\left( T_{\lambda},\cal{H} \right)$ be a irreducible unitary representation 
of $W$ on the Hilbert space $\cal{H}$ (in the next section we are going
to review a bit of representation theory of $W$). Given an element 
$(t,\vec{\alpha})$ of $W$, we denote the action on a given vector 
$\mid v \rangle$ by $T_{\lambda}(t,\vec{\alpha})\mid v \rangle$.

Coherent states are vectors $\mid \vec{\alpha} \rangle $ of $\cal{H}$, generated by the action of 
elements of $W$ in the form $(0,\vec{\alpha})$ on a fixed vector $\mid\Psi_0 \rangle$ of $\cal{H}$,
and the set of such a vectors form a coherent state system (C.S.S.). In the case that the 
vector $\mid\Psi_0 \rangle$ is a vacuum state $\mid 0 \rangle$, we call the system by a standard 
coherent state system. There are other alternative but equivalent
definitions of C.S.S. (see \cite{Per} and \cite{Kla}).

Since, we've assumed the representation $T_{\lambda}$ is irreducible, the set 
$\left\{ \mid \vec{\alpha} \rangle \right\}$ generates the whole space $\cal{H}$. Actually, the 
system $\left\{ \mid \vec{\alpha} \rangle \right\}$ is called an overcomplete system, what means 
we don't have mutual orthogonality between all the vectors of the set:\\
\begin{eqnarray}
\langle \vec \alpha \mid \vec \beta \rangle = \langle 0\mid T_{\lambda}(0,\vec{\alpha^{+}})T_{\lambda}(0,\vec{\beta}) 
\mid 0 \rangle =\exp \left( iIm(\beta \bar{\alpha} \right) \langle 0\mid T_{\lambda}(0,\vec{\beta}-\vec{\alpha})\mid 0 \rangle \\  \label{5}
\mid \langle \vec \alpha \mid \vec \beta \rangle \mid^{2} =\mid \langle 0\mid T_{\lambda}(0,\vec{\beta}-\vec{\alpha})\mid0 \rangle\mid^{2}=
\rho(\vec{\beta}-\vec{\alpha}) \label{6}
\end{eqnarray}

In fact, $\rho (\vec{\alpha} - \vec{\beta})$ can be not identically zero.

In terms of projectors, we have that the projectors
$\mid\alpha\rangle\langle
\alpha\mid$ are not mutually orthogonal projectors.

To find a orthogonal basis for the C.S.S. von-Neumann announced the 
existence of a countable orthogonal basis $\{\mid\alpha_k\rangle\}$ within
$\{\mid\alpha\rangle\}$ when we consider a lattice in the $\alpha$-plane $V$.

Let us take two non-colinear vectors $\{w_1,w_2\}$ such that $B(w_1,w_2)=2iIm(w_1\bar{w_2})\neq0$
and let us consider vectors in the form $\alpha_m=m_1w_1+m_2w_2$ / $m_1,m_2\in\bf{Z}$.

von Neumann stated that:\\
\begin{description}
\item{i)}The system $\left\{\mid\alpha_m\rangle\right\}$ is going to be over complete if 
$Im(w_1\bar{w_2}) < \pi$, and it remains over complete even if we remove a finite number 
of vectors from $\{\mid\alpha_m\ \rangle\}$\\
\item{ii)}for $Im(w_1\bar{w_2})>\pi$, the system is not complete,\\
\item{iii)}for $Im(w_1\bar{w_2})=\pi$ the system is complete and remains complete
even if we remove a single vector from $\{\mid\alpha_m\rangle\}$.\\
\end{description}

Now we have two remarks: First, since $Im(w_1\bar{w_2})$ is two times the
area of the triangle of vertices $(0,w_1,w_1+w_2)$, when we consider the 
vectors $\left\{\mid\alpha_m \rangle \right\}$ with $Im\ ( w_1\bar{w_{2}}\ )=\pi$,
 we are considering one state by cell of area $Im(w_1\bar{w_2})$, that means 
(in our normalization), one state by Planck cell (see \cite{Per} and
\cite{Kla}).

Second, we can express $exp\left(iIm(w_1\bar{w_2})\right)$ as $T_{\lambda}(w_1)T_{\lambda}(w_2)
T_{\lambda}(w_1+w_2)^{-1}=exp\left(iIm(w_1\bar{w_2})\right)$. In geometric quantization (section 4)
approach, this term is a holonomy term (see fig.1) of a holomorphic line 
bundle that make the geometric quantization of the system.

So, asserting that $exp\left(iIm(w_1\bar{w_2})\right)=\pi$, is equivalent to impose the
Bohr first quantization for the system (see for instance \cite{Woo}).

From this two remarks we wish to investigate the meaning of the overcompleteness of the C.S.S..

\section{Representations of Weyl-Heisenberg group on the lattice}

Let us consider a complete C.S.S., given by a "complete" lattice 
\begin{equation}
L=\left\{\alpha_m=m_1w_1+m_2w_2/Im(w_1\bar{w_2})=\pi,m_1,m_2\in\bf{Z}\right\}  \label{7}
\end{equation}

If we consider in $V$ the alternating bilinear form $B:V \times V \rightarrow {\bf R},
B(v,w)=Im(v \bar{w})$, we observe that, if we restrict $B$ to $L\times L$, the image of $B$ 
lies in $\pi\bf{Z}$ and for no other vector $v \in V, v \not\in L$,
we can have $B(v ,\alpha_m)\in\pi\bf{Z}$ for any vector $\alpha_m\in L$.

For a non-complete lattice: 
\begin{equation}
L=\left\{\alpha_m = m_1w_1+m_2w_2/Im(w_1\bar{w_2})>\pi,m_1,m_2\in \bf{Z} \right\},    \label{8}
\end{equation}
if for instance we consider $Im(w_1\bar{w_2})=k\pi$ for $k$ a positive integer, $k\neq1$
, we have a set of vectors $\left\{ v_{m}={m_1}/{k}w_1+ {m_2}/{k}w_2 , m_1,m_2\in{\bf Z}(modk{\bf Z})\right\}$,
such that $B(v_m,\alpha_m)\hookrightarrow \pi\bf{Z}$ for any $\alpha_m\in L$.

In the first case we say that the lattice is self-dual, and in the second case
we consider a lattice $L'=L\cup\{v_{m}\}$ dual to $L$, such that $B:L'\times L\hookrightarrow \pi{\bf Z}$.

We can describe also $L'$ as lattice generated by $\left\{ w'_1={w_1}/{k},w'_2={w_2}/{k} \right\}$  
such that: 
\begin{equation}
L'=\left\{m_1w'_1+m_2w'_2/Im(w'_1\bar{w'_2})=\frac{\pi}{k^2}(<\pi),m_1,m_2\in{\bf Z}\right\}.  \label{9}
\end{equation}

So, the dual lattice of a "non-complete" lattice is an "overcomplete lattice related to a overcomplete system.

We are going to focus now representations of the Weyl-Heisenberg group induced by characters on
the lattice.

A character of a Lie Group is a continuous complex valued function $\chi$ on $G$
such that $\mid \chi (g) \mid =1$, and $\chi (gg')=\chi(g) \chi(g')$ for $g,g'$
in $G$. The associated infinitesimal character is the linear form $\chi$ 
in the Lie Algebra $LieG$ of $G$ characterized by
$\chi\left(\exp(A)\right)=\exp\left(\chi(A)\right)$.

 Let $\chi$ be a character of some closed subgroup $H$ of a group $G$. Let 
 ${\cal H_{\chi}}$ denote the Hilbert space consisting of all functions
$f$ on 
 $G$ satisfying the following conditions:\\
 \begin{description}
 \item[(a)]$f$ is Borel measurable on $G$
 \item[(b)]$f(hg)=\chi(h)f(g)$ for $g$ in $G$ and $h$ in $H$
 \item[(c)]the integral $\int_M \mid f(g) \mid^2 dg$ is finite, for $M=G/H$ 
 \end{description}

 The norm in ${\cal H_{\chi}}$ is given by:\\
$\| f \|^2=\int_M {\mid f(g) \mid}^2 dg$.

Observe that since $\mid \chi(h) \mid =1$, $\mid f\mid^2$ is constant on every coset
$Hg$, and the expression above make sense,

To every $g$ in $G$, there is associated an unitary operator $\pi_\chi (g)$
on ${\cal H}_{\chi}$by:
\begin{equation}
\left(\pi_{\chi}(g).f\right)(g')=f(gg')       \label{10}
\end{equation}
The pair $(\pi_{\chi},{\cal H}_{\chi})$ is a representation of $G$, called
representation 
induced by the character $\chi$ of $H$.

Observe that the C.S.S. is a representation of $W$ induced by characters on the center $Z$
of $W$. The characters of $Z$ are given by the formula:
\begin{equation}
\chi_{\lambda}(t,0)=\exp 
2\pi i(\lambda t)     \label{11}
\end{equation}
where $\lambda$ runs {\bf R}, and the infinitesimal character associated
to  
$\chi_{\lambda}$ is the linear form on $\tilde z$ ($=LieZ$) is given by:\\
\begin{equation}
\chi'_{\lambda}=2\pi i\lambda    \label{12}
\end{equation}
We have the classification of irreducible unitary representations of $W$ given
 by the Stone-von Neumann theorem which asserts that:\\ 
\begin{description}
\item[(a)]For every $\lambda \neq 0$, there is, up to unitary equivalence, exactly one irreducible representation  
$(\pi,{\cal H})$ satisfying \ref{10}
\item[(b)]The case $\lambda=0$ corresponds to the representations which are 
trivial on the center $Z$ of $W$. They are the one-dimensional representations
given by the characters $\chi_u$ of $W$, given by:
\begin{equation}
\chi_u(t,v)=\exp(2\pi iB(v,u))  \label{13}
\end{equation}
\end{description}

What we are going to analyze are the representations of $W$ induced by
characters 
defined on a discrete lattice of $\Gamma_L$ on $W$, $\Gamma_L=\{(t,\alpha_m)=
m_1 w_1+m_2 w_2, t,w_1,w_2\in{\bf Z} \}$, the group which elements
are
given 
by the lattice $L$ generated by $(w_1,w_2)$.

The motivation to analyze these representation is pointed on \cite{Per} in the 
analysis of completeness of the system , (in the case that
$Im(w_1\bar{w_2})=2\pi$
where looking at the expression:\\
\begin{equation}
T_{\lambda}(\vec{\alpha_m})T_{\lambda}(\vec{\alpha_{m'}})=
T_{\lambda}(\vec{\alpha_m}+\vec{\alpha_{m'}})=
T_{\lambda}(\vec{\alpha_{m'}})T_{\lambda}(\vec{\alpha_m})  \label{14}
\end{equation}
the author ask for a common eigen-distribution for the operators $\{T_{\lambda}(\vec{\alpha_m})\}$.

If we start considering the base vectors $w_1,w_2$, we should have for the 
expression of this eigen-distribution:
\begin{equation}
T_{\lambda}(w_i)\mid \Theta \rangle=\exp(\pi i \varepsilon)\mid \Theta \rangle \label{15}
\end{equation}
because of the unitarity, where $0\leq\varepsilon<2$, for $i=1,2$.

For a generic element $\vec{\alpha_m}=m_1w_1+m_2w_2$ of $L$, we should have:\\
\begin{equation}
T_{\lambda}(\vec{\alpha_m})\mid\Theta\rangle=\exp \pi i \lambda(m_1\varepsilon_1+m_2\varepsilon_2+m_1m_2)\mid\Theta\rangle  \label{16}
\end{equation}

The general form for a character in $\Gamma_L$ \cite{Car} is:\\ 

\begin{equation}
\chi_{p,F}(t,\vec{\alpha_m})=\exp(\pi i pt)\exp(\frac{1}{2}F(\vec{\alpha_m})) \label{17}
\end{equation}
where p runs the integers and the function $F(\vec{\alpha_m})$ should satisfy 
the following congruence:\\
\begin{equation}
F(v_1+v_2)=F(v_1)+F(v_2)+pB(v_1,v_2)\;\;\;\;(mod2)   \label{18}
\end{equation}
such that we have:

\begin{equation}
T_{\lambda}(\vec{\alpha_m})\mid\theta\rangle=\exp\pi i F(\alpha_m)\mid\theta\rangle   \label{19}
\end{equation}

The general result of Cartier \cite{Car} is:

Given a representation ${\cal{D}}_{L,p,F}=(\pi(W), \cal{H})$ induced by a 
character $\chi_{p,F}$ of $\Gamma_L$, this representation is irreducible if 
and only if $L$ is a self-dual lattice (a "complete" lattice), associated with a 
complete C.S.S.. In this case ${\cal{D}}_{L,p,F}$ is isomorphic to  the 
representation induced by $\chi_p$ (item (a) of the Stone-von Neumann
theorem).

If $L$ is not complete, to ever $\lambda$' in $L'$ (mod $L$),that is,
given elements of the dual lattice $L'$ modulo the lattice $L$, we have 
an operator that commutes with the induced representation ${\cal{D}}_{L,p,F}$. 
So, if we have $[L':L]=e^2$, we have $e$-operators that commute with the 
representation, a direct sum of $e$-copies of the irreducible one (when $L$
is self-dual).
\begin{equation}
{\cal{H}}=\oplus_{i=1}^{e} {\cal{H}}_i   \label{Hil}
\end{equation}
Another result asserts that the invariance equation \ref{19} has, for a given 
character $F$ and up to constant multipliers, one unique 
solution in $\cal{H}_{-\infty}$, (the dual space of $C^{\infty}$-functions 
$\cal{H}_{\infty}$) in the case that $L$ is self-dual; and, if
$[L':L]=e^2$, 
the equation has $e$ linear independent solutions, generating a $e$-dimensional 
subspace of $\cal H_{-\infty}$.

We can contemplate both results looking at the solutions of this equation 
that lie in the ring of Jacobi theta functions, when we consider the holomorphic 
representation (the Fock-Bargmann representation) of the distributions 
$\{\mid\Theta\rangle_{\varepsilon}\}$, solutions of equation \ref{19}. 

We define a complex structure $J$ in $V$, such that $J^2=-1$ and 
$B(Jv,Jv')=B(v,v')$, $B(v,Jv)\geq0$.

We consider the complexification of $V$ to $V_{\bf{C}}$ and the natural 
extension of $B$ and $J$ to their complexified version.

We have an unique hermitian form $H$ such that:
\begin{equation}
H(v,v')=B(v,Jv')+iB(v,v')   \label{20}
\end{equation}

We consider now the representation ${\cal{D}}_{L,\lambda,F}=(\pi, {\cal{L}}^2)$ 
induced by a character $\chi_{p,F}$ of $\Gamma_L$ over the ${\cal{L}}^2$-holomorphic
functions on $V_{\bf{C}}$, with respect to the Kahler potential $-\pi\lambda H$, 
such that:

\begin{equation}
(\phi,{\phi}')=\int_V e^{-\pi\lambda H (v,v)}\phi(v) \bar{{\phi}'(v)} dv  \label{21}
\end{equation}

In the case of self-dual lattices (for $\lambda=1$); the action of $W$ in 
this representation is:

\begin{equation}
(U_v \phi)(v')=e^{-\pi \left[ \frac{H(v,v)}{2}+H(v,v') \right]}\phi(v+v')   \label{22}
\end{equation}
and the invariance equation \ref{18} takes the form:

\begin{equation}
\phi(v+\lambda)=\phi(v)\exp\left\{\pi\left[ \frac{1}{2} H(\lambda,\lambda)
+H(\lambda,v)+iF(\lambda)\right] \right\}   \label{23}
\end{equation}
for $\lambda\in L$.

This equation has one solution on the ring of the theta functions.

For $[L':L]=e^2\neq1$, we will have a set of the $e$-solution of the
equation, 
$\{\Theta_m, m\in L'modL\}$. These are the theta functions with characteristic
$m$.

These solutions can be generated acting with the $A_{\lambda}$ operators 
($\lambda\in L'modL$) on $\Theta$ (the solution of \ref{19} when $[L':L]=1$). 
Since $\{A_{\lambda}\}$ commute with the 
group $\Gamma_L$, the resulting functions
are also solutions of \ref{19}.

The linear independence of the $\Theta_m$ can be verified evaluating 
$\langle\Theta_{m,L},\Theta_{m',L}\rangle \propto \delta_{L+m,L+m'}$.

More about theta functions can be seen for instance in \cite{Vil} and \cite{Kac}.

As we've said, the theta functions are the holomorphic realizations on $L^2$  
of the distributions, solutions of \ref{19}.

Each $\Theta_{m,L}$ function in going to be related to a Hilbert
space (eq. \ref{Hil} and to a single lattice of
complete type but with
the origin dislocated, since the periodicity of the 
theta functions is given by \ref{19}, and to generated all of then we have 
translated the original $\Theta$ by steps on the $L'(modL)$ lattice.

All wee considered is easily generalized to 2n-dimension (2 p-variables and 
2 q-variables)by using theta functions of many variables. In the next section we 
are going to consider this generalization. 

\section{Geometric Quantization and C.S.S.}

To have a physical picture of this result, we're going to associate to each
$\Theta_{m,L}$ function a line bundle over the torus $T=V/L$, the 
geometric (holomorphic) quantization over $T$ (see for instance \cite{Woo},
\cite{Sny} or \cite{Kon}).

We start with a 2$n$-dimensional manyfold $(M,\omega)$, the phase space of 
the classical system, and we define a complex line-bundle with connection 
$({\cal{L}},\nabla) \stackrel{\pi}{\longrightarrow}(M,\omega)$.

The wave functions $\{ \Psi\}$ are going to be sections $\sigma:M\rightarrow
\bf{C}$ on the line bundle, and the operators $\{\hat{f}\}$ over $\cal{H}$, corresponding to the 
classical quantities $\{f:M\rightarrow\ {\bf{R}} /f\in C^{\infty}(M)\}$ are 
going to be operators that act in the sections of $L$.

The connection $\nabla$ can be defined by a connection one form $\alpha$, that 
vanishes in the horizontal vector fields ($\alpha(Y)=0$, for $Y$ horizontal) 
as follow:

\begin{equation}
\nabla_X \sigma=2\pi i\sigma^{\ast}\alpha(X)\sigma   \label{24}
\end{equation}
where $\sigma^{\ast}$ is the pull-back applied to the one form $\alpha$, and
$\sigma$ is a section.

The line-bundle with connection $({\cal{L}},M)\stackrel{\pi}{\longrightarrow}(M,\omega)$ 
is a pre-quantization of $(M,\omega)$ if $d\alpha=-(2\pi\hbar)^{-1}\pi^* \omega$
, that is, the curvature $d\alpha$ is projected in the sympletic form $\omega$.

Such a line bundle do exist if and only if $(2\pi\hbar)^{-1} \omega $ define
a deRham cohomology class over $\bf{Z}$. This condition is equivalent to 
Bohr-Sommerfeld quantization \cite{Boh}.

For this, let us take a closed path in $M$, and let lift it to $\cal{L}$ by
$\Psi$. The holonomy term is given by $\exp\left({i}/{\hbar}
\oint_{\gamma}\theta\right)=\exp\left( {i}/{\hbar}\int_S \omega \right)$ 
where $\partial S=\gamma$.

So the wave function is well defined over $\gamma$ if ${i}/{\hbar}\int_S\omega$ 
is $2\pi\bf{Z}$-valued. 
 
Observe that the phase term pointed in the second remark of section 1, 
that the wave function obtain when we circuit around the triangle of 
vertex $(0,\alpha,\alpha+\beta)$ is a holonomy term, and because of this 
the von-Neumann condition to a lattice be complete is equivalent to 
the Bohr-Sommerfeld quantization.

What we want to investigate is the overcompleteness of a C.S.S. in terms 
of line bundles. This relation is easily obtained if we focus the last result 
we have obtained, the solutions of \ref{19} expressed in terms of the theta 
functions with characteristic. This is what we wish to consider from now.

In what follow where we write line bundles we are talking about holomorphic 
line bundles. For details see, for instance \cite{Gri}.
Given a open cover $\{U_{\alpha}\}$ of $M$,a line bundle $(\cal{L},M)$
can described by a collection of transition functions 
$\{g_{\alpha\beta}\in\cal{O}^*(U_{\alpha}\cap U_{\beta})\}$ that satisfy:

\begin{eqnarray}
g_{\alpha \beta}g_{\beta \alpha}=1    \label{25}\\
g_{\alpha \beta}g_{\beta \gamma}g_{\gamma \alpha}=1   \label{26}
\end{eqnarray}

Transition functions can be defined in terms of local trivializations 
$\{\varphi_{\alpha}\}$ by:
\begin{equation}
g_{\alpha \beta}=\varphi_{\alpha} \varphi_{\beta}^{-1}       \label{27}
\end{equation}

Two sets of trivializations $\{\varphi_{\alpha}\}$, 
$\{\varphi'_{\alpha}\}$
define the same line bundle if $\varphi'_{\alpha}=\varphi_{\alpha}.f_{\alpha}$ , 
$f_{\alpha} \in \cal{O}^*(U_\alpha)$.
The equations \ref{25} and \ref{26} state that $\{g_{\alpha \beta}\}$
is a $\check{C}$ech cocycle. And, by the last paragraph, two cocycles 
$\{g_{\alpha \beta}\}$, $\{g'_{\alpha \beta}\}$ give the same line
bundle if they differ by a $\check{C}$ech coboundary, that is, the set 
of line bundles on $M$ is just by $H^1 (M,{\cal{O}}^*)$.

We are going to need now certain conceptions from sheaves cohomology
sequences.

Given an exact sequence of sheaves:

\begin{equation}
...\rightarrow \bf{Z} \rightarrow \cal{O} \stackrel{\exp}{\rightarrow}
\cal{O}^* \rightarrow...      \label{28}
\end{equation}
we have the long sheaves cohomology exact sequence:

\begin{equation}
...H^1(M,{\bf Z}) \rightarrow H^1(M,{\cal {O}}) \rightarrow H^1(M,{\cal{O}}^*)
\rightarrow H^2(M,{\bf Z}) \rightarrow H^2(M,\cal{O}) ...   \label{29}
\end{equation}

We have the boundary map in the cohomology:

\begin{equation}
H^1(M,{\cal{O}}^*) \stackrel{\delta}{\rightarrow} H^2(M,\bf{Z})    \label{30}
\end{equation}

The image $\delta({\cal{L}})=c^1 ({\cal{L}})$ of $\cal{L}$ in $H^2(M,\bf{Z})$
is the (first) Chern class of $\cal{L}$. We have that: 

\begin{equation}
c^1({\cal{L}})=\left[ \frac{i}{2\pi} \Theta \right] \in
H^2_{DR}(M)   \label{31}
\end{equation}
where $\Theta$ is the curvature of $\cal{L}$ and $H^2_{DR}(M)$ is the
second deRham cohomology group (for details see \cite{Gri}).

Consequently with this result and the first part of this
section we see that each Chern class $c^1(\cal{L})$ gives a quantization
of the system. 

In the case that $M=T=V/L$, we are going to see how theta function are
related to all these facts.

Let ${\cal{L}} \rightarrow M=V/L$ be a line bundle over the complex torus
$M$, and let $\pi^* \cal{L}$ be the pullback of $\cal{L}$ to $V$. Since any
line bundle over $V$ is trivial we can find global trivializations:

\begin{equation}
\varphi:\pi ^* {\cal{L}} \rightarrow V \times \bf{C}        \label{32}
\end{equation}
For $z \in V$, $\lambda \in L$, the fibers of $\pi^*$ at $z$ and
$z+\lambda$ are both identified with the fiber of $\cal{L}$  at $\pi(Z)$,
and comparing the trivialization $\varphi$ at $z$ and $z+\lambda$ we have
automorphism of $\bf{C}$, given as multiplication by a nonzero
complex number $e_{\lambda}(z)$, and we obtain a collection of
functions:
\begin{equation}
\{e_{\lambda}\in {\cal{O}}^* (V)\}_{\lambda \in L}      \label{33}
\end{equation}
called set of multipliers for $\cal{L}$.

This functions $e_{\lambda}$ satisfy a compatibility relation:\\
\begin{equation}
e_{\lambda '}(z+\lambda)e_{\lambda}(z)=e_{\lambda}(z+\lambda ')e_{\lambda
'}(z)=e_{\lambda+\lambda '}(z)                        \label{34}
\end{equation}
for all $\lambda, \lambda' \in L$.

It's possible to show that any line bundle $({\cal{L}},M)$ can be
given by a set of multipliers $\{e_{\lambda}(z)\}$ and that up to a
translation in $M$ all line bundles is determined by its Chern class.

In the prove of these results we fix the multipliers to be:

\begin{equation}
e_{\lambda_{\alpha}}(z)=1,\hspace{1,5cm} e_{\lambda_{n+\alpha}}(z)=e^{-2\pi 
iz_{\alpha}} \hspace{1,5cm}\alpha=1,...,n \label{35}
\end{equation}

For a basis $\lambda_1 ,....,\lambda_{2n}$ for $L$ and a dual system of
coordinates such that the curvature $\omega$ is given by:\\
\begin{equation}
\omega=\displaystyle\sum_{\alpha=1}^n \delta_{\alpha}dx_{\alpha} 
\wedge dx_{n+\alpha}     \label{36}
\end{equation}
where $e_{\alpha}=\delta_{\alpha}^{-1}\lambda_{\alpha}$, $\alpha=1,...,n$.

The Chern class in this trivialization will be $c_1(\cal{L})=[\omega]$.

Now we want to consider the set of line bundles having a given positive the same 
Chern class. For any $\mu \in M$, the translation $\tau_{\mu}:M \rightarrow M$ is 
homotopic to the identity and hence for any line bundle $({\cal{L}},M)$:\\

\begin{equation}
c_1 (\tau_{\mu}^* {\cal{L}})=c_1 (\cal{L})        \label{37}
\end{equation}

Actually it's possible to prove that any line bundle having the same Chern class 
as $\cal{L}$ must be a translate of $\cal{L}$.

If the multipliers of $\cal{L}$ is like \ref{35}, the set of multipliers of 
$\cal{L}'=\tau_{\mu}^*M$ is going to be:

\begin{eqnarray}
e'_{\lambda_{\alpha}}(z) &=& e_{\lambda_{\alpha}}(z+\mu)\equiv 1    \label{38}\\
e'_{\lambda_{n+\alpha}}(z)&=&e_{\lambda_{n+\alpha}}(z+\mu) \nonumber \\
 &=& e^{-2\pi i(z_{\alpha}+\mu_{\alpha})}    \label{39}
\end{eqnarray}

Now, for a given section $\tilde{\theta}$ of ${\cal{L}}$ over $U\subset M$,
$\theta=\varphi^*(\pi^*\tilde{\theta})$ is an analytic function of
$\pi^{-1}(U)$ satisfying:\\
\begin{eqnarray}
\theta(z+\lambda_{\alpha})=\theta(z)    \label{40}\\
\theta(z+\lambda_{n+\alpha})=e^{-2\pi iz_{\alpha}}\theta(z)    \label{41}
\end{eqnarray}
and conversely any such function defines a section of $\cal{L}$.

Now for $\mu=\frac{1}{2} \sum Z_{\alpha \alpha}.e_{\alpha}$, let ${\cal{L}}'=\tau_{\mu}^* {\cal{L}}$, with the multipliers given by equations \ref{38} and \ref{39}.

If $\tilde{\theta}'$ are global sections of $\cal{L}'$, we will have holomorphic functions 
on $V$, just like the functions given in equations \ref{40} and \ref{41}:

\begin{eqnarray}
\theta'=(z+\lambda_{\alpha})=\theta'(z)     \label{42}\\
\theta(z+\lambda_{n+\alpha})=e^{-2\pi iz_{\alpha}-\pi iZ_{\alpha \alpha}} \label{43}
\end{eqnarray}

These are the equations for the theta functions. The matrix $Z_{\alpha \beta}$ constitute 
part of the so called period matrix (see \cite{Gri}).

So we see how the theta functions are related with line bundles for a given Chern class. 
To obtain the theta functions with characteristic considered in the previous section we 
just have to consider translations $\tau_{\mu}^*{\cal{L}}$ (where ${\cal {L}}$ is a line bundle with Chern class equal to one, related to the original theta function) of a fixed size, the size of the minimun cell related to $L'$, 
the dual lattice of a non-complete lattice (eq. \ref{19}). These translations is going to 
operate just like the $A_{\lambda}$ operators we mentioned in the previous section. 

Another way to see these theta functions is to consider, in the first case (complete case), 
a principally polarized complex torus (associated to a self-dual or a complete lattice eq. 
\ref{7})and the theta function is the only global section of it. In the second case, we 
consider a polarized torus (associated to a overcomplete lattice, eq. \ref{9}) with 
polarization given by the Chern class (eq. \ref{36}) $c_1 ({\cal {L}})=\Pi \delta_{\alpha}$. 
The set of theta functions is going to be the global sections of this torus. This is 
equivalent to consider the overcomplete lattice itself and to associate with each theta 
function with characteristic a complete lattice belonging to the overcomplete one. 

More about this can be seen in \cite{Gri} and \cite{Igu}.

\section{Application and Remarks}

We have seen how we can associate certain types of overcomplete C.S.S. to 
a set of quantizations of the classical system. To have a more concrete picture 
of this scenario, let us consider the hamiltonian $H_0 = \frac{1}{2}({\hat p}
^2 + {\hat{q}}^2 - \hbar)$ and let $\mid \Psi_0 \rangle$ be the ground state of $H_0$.

If we map $\mid \Psi_0 \rangle$ in a coherent state $\mid \alpha \rangle=D(\alpha
) \mid \Psi_0 \rangle$, this coherent state is going to be the ground state of 
the conjugated hamiltonian $H_{\alpha}=D(\alpha)H_0 D(\alpha)^{-1}$ of $H_0$, 
that is:
\begin{equation}
H_{\alpha}\mid \alpha \rangle=0    \label{44}
\end{equation}

If we have translation invariance in the problem we could have these
coherent states 
representing degenerated states.

This is the case in Landau levels. In \cite{Kla} we have a phase space approach
to the problem where the coherent states are used as a basis for the
propagator kernel. In
this work the authors, using the Riemann-Roch theorem, obtain the
degeneracy of the 
lowest Landau level $n+1-g$ where
$n$ is an integer number expressing the normalized magnetic charge plus the Euler
characteristic of the surface, and $g$ the genus of the surface. This result has been 
already obtained without mention of coherent states by pure geometrical arguments 
in \cite {Mar}.

In the present work the Riemann Roch theorem can be used directly. First
we have to observe that if
we consider a polarized torus originated by the ``overcomplete'' lattice,
that is, with a 
minimun cell with area less then $\pi$ (eq. \ref{9}), the Riemann Roch theorem gives (see 
\cite{Gri}):
\begin{equation}
dim H^0 (M,{\cal{O}}({\cal{L}}))=\Pi_{\alpha} \delta_{\alpha},
\end{equation}
that is exactly the number of theta functions with characteristic or equivalently the 
number of Hilbert spaces in the direct summation \ref{Hil}.
\vspace{3cm}

{\bf Acknowledgment}
The author is grateful to IFT-UNESP, IMECC-UNICAMP and Inst. of Maths.-TCD  for the
 conditions to develop this research. The author is very grateful also to
Prof. 
S. Sen in TCD, Prof. M.A.F.Rosa, Prof. L. San Martin, Prof. A. Ananin in Inst. of 
Maths.-IMECC-UNICAMP for the very precious elucidations and discussions. 
A very special thanks to A. Iorio for has presented me the theme, for the 
very passionate and rich discussions about this and many other themes of
physics
 and life. Finally I thank God for everything this work has involved.     
This work was suported by Conselho Nacional de Pesquisa (CNPq) and Funda\c c\~ao de Amparo \`a Pesquisa do Estado de S\~ao Paulo - (FAPESP).


\begin{thebibliography}{99}

\bibitem{Kla} R.Alicki, J.R. Klauder, J. Lewadowski, {\it Landau Level 
Ground State Degeneracy, and its Relevance for a General Quantization 
Procedure}, Pre-print gr-qc/9312006
\bibitem{Boh} D. Bohn, {\it, Quantum Theory}, Prentice hall Inc.,
Englewood Cliffs, 1951
\bibitem{Car} P. Cartier, in {\it Proc. Symp. Pure Math.} vol 9,
Algebraic Groups and Discontinu
ous Groups, Am. Math. Soc., Providence,
1996
\bibitem{Gri} P. Griffiths, J. Harris, {\it Principles of Algebraic
Geometry}, John Wiley \& Sons, New York, 1978
\bibitem{Igu} J. Igusa, {\it Theta functions}, Springer Verlag, Berlin, 1972.
\bibitem{Kac} V. G. Ka${\check c}$, D.H. Peterson, {\it Inf. Dim. Lie
Algebra, Theta functions and Modular Forms}, in Advances in Maths.,
vol.53, n.2, 1984
\bibitem{Kla} J.R. Klauder, Bo-S. Skagertan, {\it Coherent States -
Applications in Physics and Mathematical Physics}, World Scientific,
Singapore, 1985
\bibitem{Kon} B. Konstant, {\it Quantization and Unitary Representations},
in Lectures in Modern Analysis and Applications III, Lecture Notes in
Math., vol170, 1970
\bibitem{Mac} G. W. Mackey, {\it Induced Representation of Groups and
Quantum Mechanics}, W.A. Benjamin, Inc. New York, and Editor Borighieri 
Torino, 1968
\bibitem{Mar} P. Maraner, {\it Landau Ground State on Riemannian Surfaces},
Mod. Phys. Letters A, vol.7, no.27, 2555-2558, 1992
\bibitem{Per} A. Perelomov, {\it Generalized Coherent States and Their
Applications}, Springer Verlag, Berlin, 1986
\bibitem{Sat} H. Sato, {\it Landau Levels and Quantum Groups}, Mod. Phys.
Letters A, vol.9, No.5, 451-458, 1994.
\bibitem{Shi} A.Shimabukuro {\it Geometrical and Topological Aspects of 
Quantization of Field Theories}, Phd thesis, IMECC-UNICAMP, Campinas,1998.
\bibitem{Shi2} A. Shimabukuro, In preparation
\bibitem{Sny} J. Sniatycki, {\it Geometric Quantization and Quantum
mechanics}, Springer Verlag, New York, 1980
\bibitem{Vil} N. Ja. Vilenkin, A.U. Klimyk, {\it Representation of Lie
Groups and Special Functions}, vol. 1,2 \& 3, Kluwer Academic Publishers,
Dordrecht, 1992
\bibitem{Woo} N.Woodhouse, {\it Geometric Quantization}, Oxford University Press,
Oxford, 1980

\end{thebibliography}
\end{document}